\documentstyle[aps,preprint]{revtex}
\newcommand{\be}{\begin{equation}}
\newcommand{\ee}{\end{equation}}
\newcommand{\bea}{\begin{eqnarray}}
\newcommand{\eea}{\end{eqnarray}}
\begin{document}
\bibliographystyle{prsty}
\title{Attractive Interaction Between Pulses in a Model for Binary-Mixture
Convection}
\author{Hermann Riecke}
\address{
Department of Engineering Sciences and Applied Mathematics \\
Northwestern University, Evanston, IL 60208, USA
}
\maketitle
\begin{abstract}

Recent experiments on convection in binary mixtures have shown
that the interaction between localized waves (pulses) can be repulsive as well
as
{\it attractive} and depends strongly
on the relative {\it orientation} of the pulses.
It is demonstrated that the concentration mode, which is characteristic
of the extended Ginzburg-Landau equations introduced recently, allows a natural
understanding of that result. Within the standard complex Ginzburg-Landau
equation this
would not be possible. 


\noindent

\end{abstract}


Convection in binary mixtures exhibits an extremely richness of
dynamics. Among the most striking features are localized wave packets which
form
stable, particle-like entities drifting through the system. They have been
studied
experimentally in great detail resulting in a
large body of data \cite{KoBe88,BeKo90,NiAh90,MoFi87,Ko91,Ko91a,Ko93,Ko94}.
On the theoretical side these results have raised a number
of questions  regarding
their shape, regime of existence, stability, and drift velocity.
They have been addressed by a number of authors employing full numerical
simulations of
the Navier-Stokes equations \cite{BaLu91,BaLu94} as well as analytical and
numerical
studies of the
complex Ginzburg-Landau equation (CGL) \cite{ThFa88,Pi87,MaNe90,HaPo91}. The
simulations
of the Navier-Stokes equations were able to reproduce various features of the
experimental
pulses and gave detailed insight into the concentration field which causes the
oscillatory dynamics of the convection and appears to be instrumental in
stabilizing
the pulses. The studies of the CGL show that within
that framework the localization of the waves is due to dispersion.

It has been pointed out, however, that various important qualitative
features of the pulses cannot be captured within the CGL.
Among them is the anomalously slow drift of the pulses and their stability
behavior.
In previous work I suggested that this is due to the relevance of
a concentration mode which becomes an independent dynamical variable for the
small
mass diffusion in the liquids employed \cite{Ri92}. Based on this observation I
derived a set of extended Ginzburg-Landau equations (ECGL) which are
characterized by
an additional field, a slow concentration mode. Various analytical and
numerical
analyses showed that this extension is able to account for the
difficulties discussed above \cite{Ri92,Ri92a,HeRi95,Ri95,RaRiprep}. In the
present
communication I discuss an additional difficulty which has not been
addressed previously.

An interesting set of experiments deals with the interaction of pulses
in head-on and tail-on collisions \cite{Ko91a}. It has been found that both
types of
collisions can result in stable bound pairs of pulses. In these experiments use
is made of
the fact that the drift velocity depends on the Rayleigh number and can in fact
become
opposite to the phase velocity (`backward' drifting pulse).
To discuss the experimental results it is useful to call the `head' of a pulse
that side
towards which the waves inside the pulse travel. There are then two types of
 pairs consisting of couterpropagating pulses:
those touching with their heads (`HH') and those touching with their tails
(`TT'). In the
experiments both
types were found to be stable over a range of parameters. The HH-pairs are only
stable if the
drift velocity of the pulses pushes them together. Thus the pulses in such a
pair
 experience only a repulsive
interaction. In the TT-pairs, however, the pulses exhibit a repulsive
interaction at small distances and
a very weak {\it attractive interaction} at larger distances. The pulses
can therefore be bound together even
if their drift velocity tends to pull them apart.

Within the conventional complex Ginzburg-Landau equations the pulse interaction
arises from
a complex cubic cross-coupling term ($g$, see below). Its real part induces a
renormalization of the growth rate of one wave by the other. Its imaginary part
affects the frequency
(cross-phase-modulation).
In the regime of interest, the real part is negative,
i.e. counterpropagaging waves suppress each other.
For pulses this leads naturally to a repulsive interaction. The imaginary part
also contributes to the interaction. In the limit of vanishing dissipation,
i.e. for the nonlinear Schr\"odinger equation,
 it has been shown that it
can lead to stably bound pulse pairs if it has the correct sign
\cite{Ma92,UeKa90}.
In a two-dimensional analysis, however, the opposite sign
is found for the regime in question \cite{ScZi93}\footnote{In fact,
this is also true for the coefficient $c_i$ (see below)
of the self-phase-modulation. A soliton perturbation theory appears therefore
not
adequate for this regime.}. Moreover, since the phase velocity,
which defines the head and the tail of the
pulse, does not enter the CGL, both HH- and TT-pairs behave identically within
that framework.

In the present communication I study the extended Ginzburg-Landau
equations (ECGL) introduced earlier \cite{Ri92} and focus on the interaction
between
pulse pairs. Using numerical simulations I show that the additional
concentration mode
characteristic of these equations can lead to an attractive interaction as
observed in the
experiments \cite{Ko91a}: the enhanced
local growth rate of the convective mode behind a pulse can act as `glue'
within a TT-pair.
For the HH-pair, however, this mechanism does not exist.

As discussed previously the slow mass diffusion in liquids necessitates
the introduction of an additional (concentration) mode even for quite small
convective amplitudes \cite{Ri92,Ri92a}. A minimal model describing the effect
of the
concentration mode on the buoyancy of the liquid \cite{BaLu91,BaLu94} is given
by \cite{Ri92}
\bea
\partial_t A + s \partial_x A& = & d \partial_x^2 A + (a +fC) A + c|A|^2A +
p|A|^4 A
+ g |B|^2A +...\label{e:ecglA}\\
\partial_t B - s \partial_x B& = & d^* \partial_x^2 B + (a^* +f^*C) B +
c^*|B|^2B + p^*|B|^4 B
+g^* |A|^2 B +...\label{e:ecglB}\\
\partial_t C & = &\delta \partial_x^2 C - \alpha C + h_2 \partial_x
(|A|^2-|B|^2)
 +...\label{e:ecglC}
\eea
The complex
amplitude of right- and left-traveling waves is given by $A$ and $B$,
respectively.
The additional, real concentration mode $C$ satisfies a diffusion equation with
damping and
is advected by the waves. In general all coefficients in
(\ref{e:ecglA},\ref{e:ecglB})
except for the group velocity $s$
are complex. Here I will, however, consider the simplified case in which $C$
affects
only the growth rate of $A$ and $B$, i.e. $f$ is assumed real as well. The
general case
has been studied for short pulses in \cite{Ri95}. The localization of long
pulses
by the concentration mode has been discussed at length in \cite{HeRi95}. The
cross-over
between long and short pulses is addressed in \cite{RaRiprep}.

\begin{figure}[htb]
\begin{picture}(420,250)(0,0)
\put(0,-45) {\includegraphics{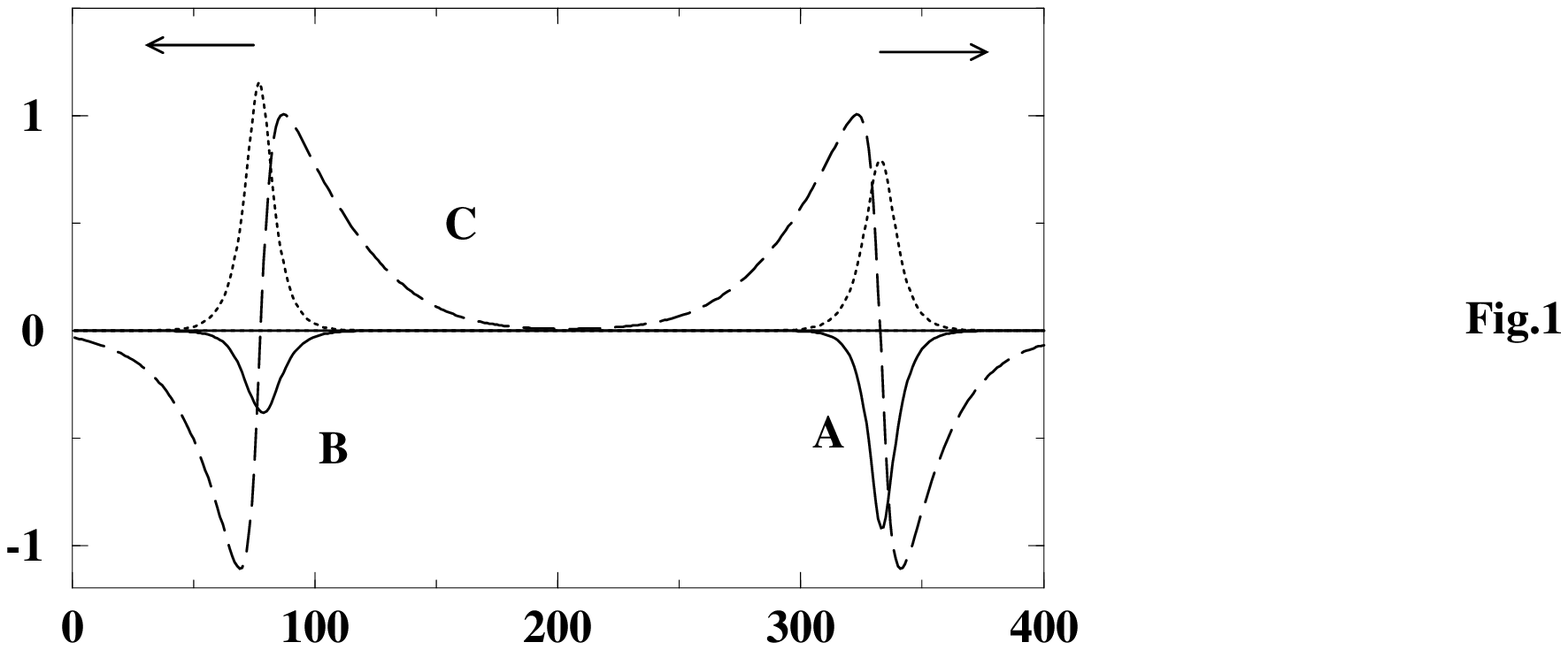}}
\end{picture}
\caption{Typical pair of widely separated pulses. The parameters are given by
$d=0.15+i$, $a=-0.24$, $f=0.4$, $c=2.4+2i$,
$g=-10$, $p=-1.65+2i$, $\alpha=0.02$, $\delta=0.25$, $h_2=0.5$.
\protect{\label{f:paarweit}}
}
\end{figure}

Fig.\ref{f:paarweit} shows a typical situation of two widely separated,
essentially
independent pulses. Their tails show the characteristic positive $C$-field
which enhances the growth rate of the respective convective amplitudes ($f>0$).
As discussed
 in detail previously \cite{HeRi95,Ri95} this can slow down the pulse. It is
reasonable to  expect  that
for sufficiently small distances between the pulses each pulse may not only
feel the
contribution to $C$ from itself but also from the other pulse. This would lead
to a
further slow-down and could amount to an attractive interaction.

\begin{figure}[htb]
\begin{picture}(420,250)(0,0)
\put(0,-45) {\includegraphics{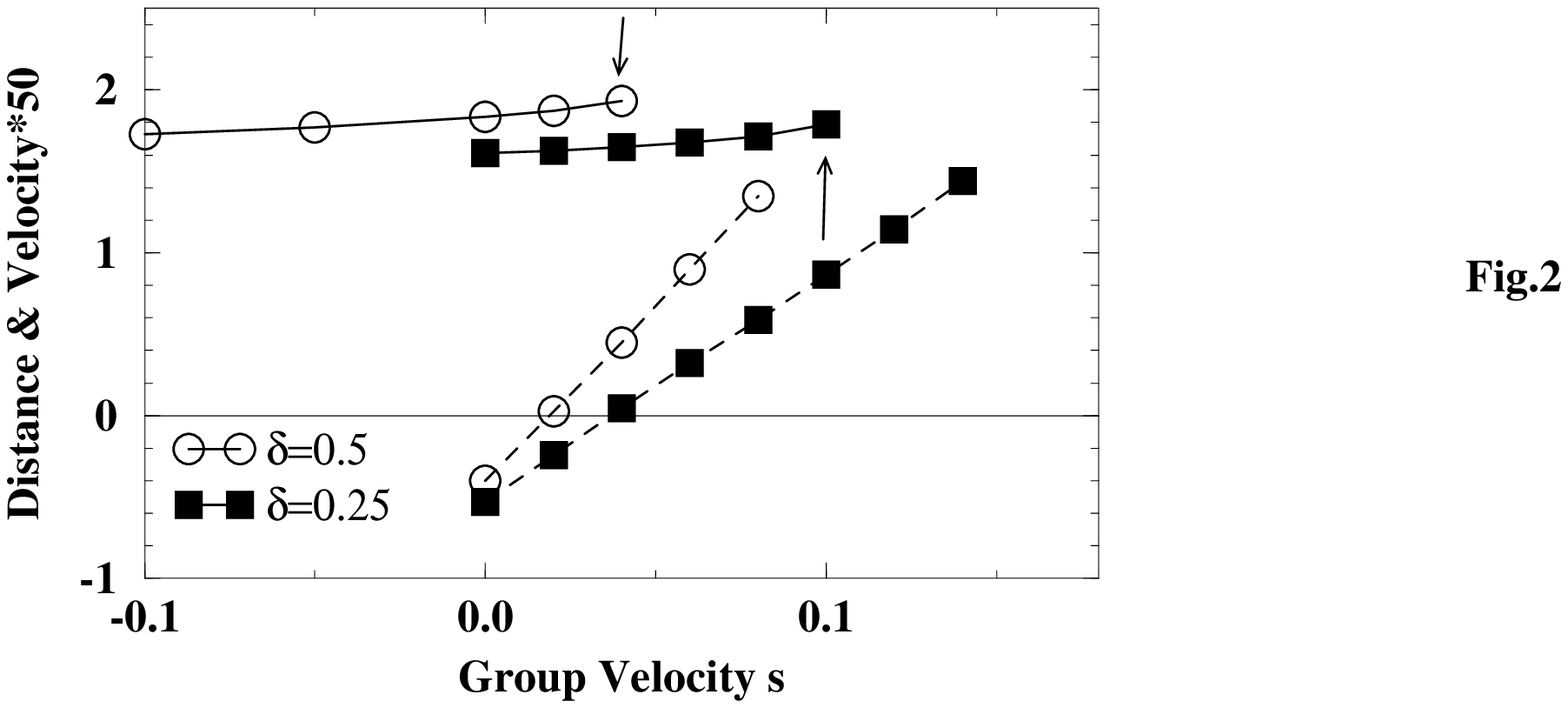}}
\end{picture}
\caption{ Dependence of the distance $D$ between the pulses and of the drift
velocity $v$
of a single pulse as a function of the group velocity $s$. Parameters are as in
fig.\protect{\ref{f:paarweit}}.
\protect{\label{f:paarLs}}
}
\end{figure}

To investigate the possibility of such an attractive interaction
eqs.(\ref{e:ecglA})-(\ref{e:ecglC}) are solved numerically
 with periodic boundary conditions. Two pulses are placed in the system and the
group
velocity is chosen such that they collide with each other in a
TT-configuration,
i.e. $s$ is chosen negative at first.
Once they have reached a steady state the equilibrium distance
is measured as a function of the group velocity.
The results are shown in fig.\ref{f:paarLs} for $d=0.15+i$, $a=-0.24$, $f=0.4$,
$c=2.4+2i$,
$g=-10$, $p=-1.65+2i$, $\alpha=0.02$, $h_2=0.5$.
The solid lines give the distance $D$ between the pulses for two values of the
diffusion
coefficient $\delta$ of the concentration mode.
The arrows indicate the largest value $s_{max}$ for which  a stable bound pair
is
still obtained.
If $s$ is increased beyond $s_{max}$ the pulses separate from each other and
$D$ diverges.
 The  crucial test for the existence of an attractive interaction is a
comparison of
$s_{max}$ with the value $s_0$ for which the
 drift velocity $v$ of a single pulse becomes positive, i.e. the value $s_0$
for which
the pulses would separate in the absence of any interaction.
 The numerically determined velocity
of   single pulses is given by the dashed lines. Clearly, it becomes positive
well before $s_{max}$. Thus, there is a finite range of parameters between
$s_0$ and $s_{max}$
in which
the concentration mode  acts as a `glue' between pulses which pulses the pulses
together.

\begin{figure}[htb]
\begin{picture}(420,250)(0,0)
\put(0,-45) {\includegraphics{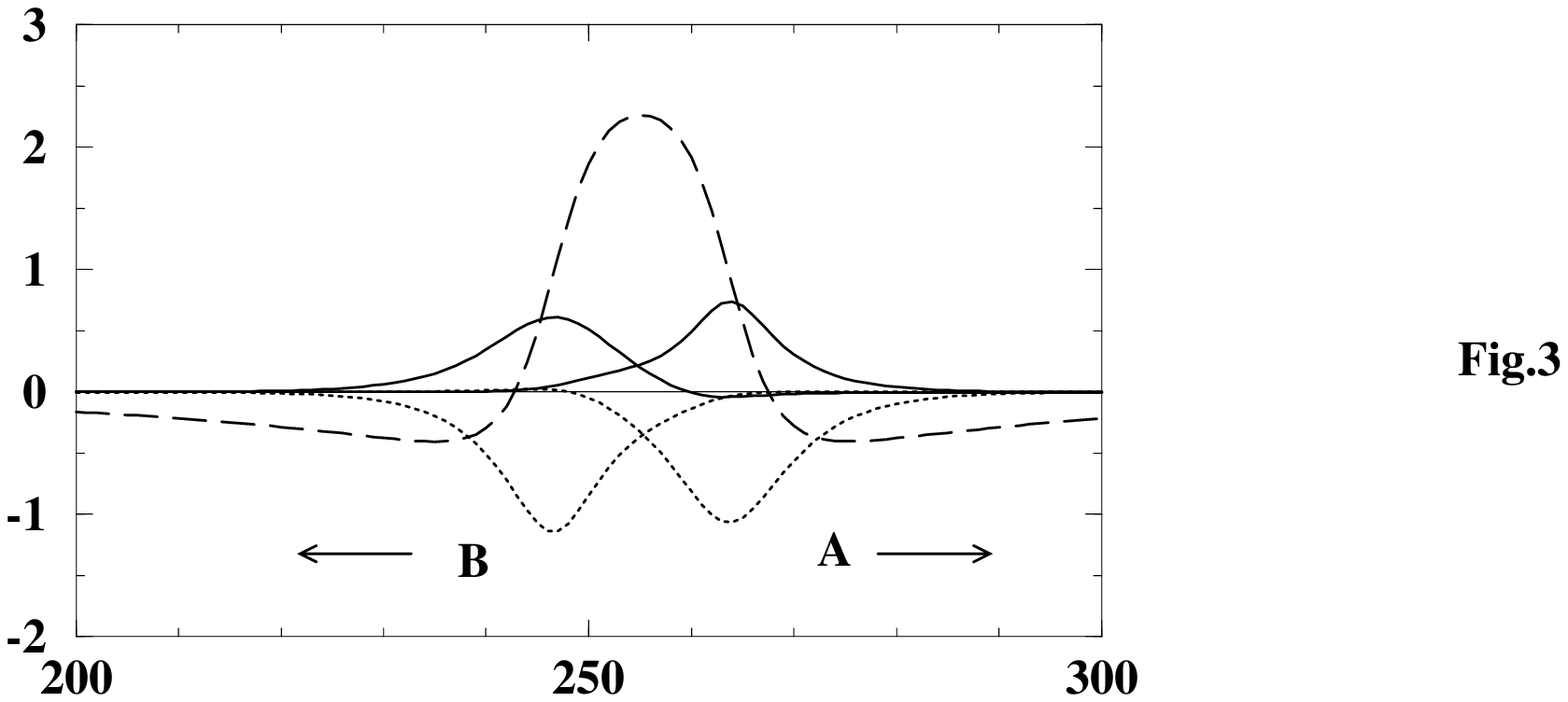}}
\end{picture}
\caption{Typical pair of stably bound pulses. The local growth rate is greatly
enhanced
between the pulses due to the accumulation of the concentration mode $C$. This
effectively `glues' the pulses together. Parameters are as in
fig.\protect{\ref{f:paarweit}}.
\protect{\label{f:paareng}}
}
\end{figure}

A typical `glued' pulse
pair is shown in fig.\ref{f:paareng}. It is characterized by a
$C$-field which is strongly enhanced between the pulses as compared to the
widely
separated case shown in fig.\ref{f:paarweit} (Note the  scale for $C$ differs
in the
two figures). No such effect is possible for HH-pairs.
To the contrary,  there the concentration mode enhances the repulsive
interaction
which is already present due to
the suppressing cross-coupling term proportional to $g$ in
(\ref{e:ecglA},\ref{e:ecglB}).
Thus, within the ECGL a clear distinction exists between HH-pairs and pairs of
the
TT-type.

In conclusion, I have demonstrated that the attractive interaction
between certain pulses observed experimentally \cite{Ko91a} can be explained
naturally if the slow concentration mode is taken into account. It also
accounts for
the observed dependence of the interaction on the relative orientation of the
pulses. The standard complex Ginzburg-Landau equation does not capture these
phenomena.

It would be interesting to study also the dynamics of collisions of pulses
within
the extended Ginzburg-Landau equations and compare them with results for the
standard
equation \cite{BrDe89a}. In the experiments a drastic collapse of the TT-pair
is
reported when the pulses are pushed together too strongly \cite{Ko91a}.

I gratefully acknowledge discussions with P. Kolodner and M. L\"ucke.
This work was supported by DOE through grant (DE-FG02-92ER14303) and by
an equipment grant from NSF (DMS-9304397).


\begin{thebibliography}{10}

\bibitem{KoBe88}
P. Kolodner, D. Bensimon, and C. Surko, Phys.~Rev.~Lett. {\bf 60},  1723
  (1988).

\bibitem{BeKo90}
D. Bensimon, P. Kolodner, and C. Surko, J. Fluid Mech. {\bf 217},  441  (1990).

\bibitem{NiAh90}
J. Niemela, G. Ahlers, and D. Cannell, Phys.~Rev.~Lett. {\bf 64},  1365
  (1990).

\bibitem{MoFi87}
E. Moses, J. Fineberg, and V. Steinberg, Phys. Rev. A {\bf 35},  2757  (1987).

\bibitem{Ko91}
P. Kolodner, Phys.~Rev.~Lett. {\bf 66},  1165  (1991).

\bibitem{Ko91a}
P. Kolodner, Phys.~Rev. A {\bf 44},  6466  (1991).

\bibitem{Ko93}
P. Kolodner, Phys. Rev. E {\bf 48},  4187  (1993).

\bibitem{Ko94}
P. Kolodner, Phys. Rev. E {\bf 50},  2731  (1994).

\bibitem{BaLu91}
W. Barten, M. L\"ucke, and M. Kamps, Phys. Rev. Lett. {\bf 66},  2621  (1991).

\bibitem{BaLu94}
W. Barten, M. L\"ucke, M. Kamps, and R. Schmitz, preprint  (1994).

\bibitem{ThFa88}
O. Thual and S. Fauve, J. Phys. France {\bf 49},  1829  (1988).

\bibitem{Pi87}
L. Pismen, Phys.~Rev. A {\bf 35},  1873  (1987).

\bibitem{MaNe90}
B. Malomed and A. Nepomnyashchy, Phys.~Rev. A {\bf 42},  6009  (1990).

\bibitem{HaPo91}
V. Hakim and Y. Pomeau, Eur. J. Mech. B Suppl {\bf 10},  137  (1991).

\bibitem{Ri92}
H. Riecke, Phys. Rev. Lett. {\bf 68},  301  (1992).

\bibitem{Ri92a}
H. Riecke, Physica D {\bf 61},  253  (1992).

\bibitem{HeRi95}
H. Herrero and H. Riecke, Physica D  (in print).

\bibitem{Ri95}
H. Riecke, preprint  (1995).

\bibitem{RaRiprep}
W.-J. Rappel and H. Riecke,   (in preparation).

\bibitem{Ma92}
B. Malomed, Phys.~Rev. A {\bf 45},  8321  (1992).

\bibitem{UeKa90}
T. Ueda and W. Kath, Phys. Rev. A {\bf 42},  563  (1990).

\bibitem{ScZi93}
W. Sch\"opf and W. Zimmermann, Phys.~Rev.~E {\bf 47},  1739  (1993).

\bibitem{BrDe89a}
H. Brand and R. Deissler, Phys.~Rev.~Lett. {\bf 63},  2801  (1989).

\end{thebibliography}

\end{document}